\documentclass[aps,preprintnumbers,twocolumn,amsmath,amssymb,nofootinbib]{revtex4}

\usepackage{amsmath,amssymb}
\usepackage{graphicx}
\newcommand{\tc}{\chi}
\newcommand{\amp}{&\!\!}
\newcommand{\mpl}{\bar{m}_{\mbox{\tiny{Pl}}}}
\newcommand{\Tosc}{T_{\mbox{\tiny{osc}}}}
\newcommand{\Lqcd}{\Lambda_{\mbox{\tiny{Q}}}}
\newcommand{\TGH}{T_{\mbox{\tiny{GH}}}}
\newcommand{\TRH}{T_{\mbox{\tiny{max}}}}

\newcommand{\rhoi}{\rho_{\mbox{\tiny{iso}}}}
\newcommand{\M}{\Lambda}
\newcommand{\cdm}{\xi_{\mbox{\tiny{CDM}}}}
\newcommand{\faN}{f_a}
\newcommand{\fh}{\hat{f}_a}
\newcommand{\eff}{\epsilon_{\mbox{\tiny{eff}}}}

\begin{document}

\author{
Mark P. Hertzberg$^1$\footnote{Electronic address: mphertz@mit.edu},
 Max Tegmark$^1$, and Frank Wilczek$^{1,2}$}
\address{$^1$Dept.~of Physics, Massachusetts Institute of Technology, Cambridge, MA 02139, USA}
\address{$^2$Peierls Center for Physics, Oxford University, OX13NP, UK}

\title{Axion Cosmology and the Energy Scale of Inflation}


\begin{abstract}
We survey observational constraints on the parameter space of inflation and axions
and map out two allowed windows: the classic window and the inflationary anthropic window. The cosmology of the latter is particularly interesting; inflationary axion cosmology predicts the existence of isocurvature fluctuations in the CMB, with an amplitude that grows with both the energy scale of inflation and the fraction of dark matter in axions.   
Statistical arguments favor a substantial value for the latter, and so current bounds on isocurvature fluctuations imply tight constraints on inflation.   For example, an axion Peccei-Quinn scale of $10^{16}$\,GeV  excludes any inflation model with energy scale 
$> 3.8\times10^{14}$\,GeV ($r> 2\times10^{-9}$) at 95\% confidence, and so implies negligible gravitational waves from inflation, but suggests appreciable isocurvature fluctuations.
\end{abstract}
\vspace*{-\bigskipamount} \preprint{MIT-CTP-3950}

\maketitle

\section{Introduction}

Early universe inflation and the QCD axion provide explanations for otherwise mysterious features of the universe.  Here we argue that assuming both at once leads to very significant constraints on their central parameters, and to highly falsifiable predictions.  

\subsection{Energy Scale of Inflation}\label{EnInf}

Inflation is the leading paradigm for early universe phenomenology \cite{Inf1,Inf2,Inf3}.
Its mechanism and the values of its central parameters are unknown, however.   
One central parameter is the energy scale of inflation $E_I$, defined as the fourth root of the inflationary potential energy density, evaluated when the modes that re-enter the horizon today left the horizon during inflation. $E_I$ is subject to both theoretical and observational constraints,
as illustrated in Figs.~\ref{Naive} \& \ref{ConsPlot}.

A multitude of inflation models involving a broad range of energy scales have been discussed in the literature, 
including chaotic inflation \cite{Inf4}, brane inflation \cite{Tye,KKLMMT} and others \cite{LiddleLythBook}.
However, very high $E_I$ has been argued to be theoretically problematic, at least for single field slow-roll inflation, because it involves super-Planckian displacements of the inflaton field $\phi$ \cite{LythBound}.
From Ref.~\cite{Baumann} the region in which $\phi$ moves at least two Planck masses is 
$E\gtrsim2.4\times10^{16}$\,GeV. 
The intuition that high $E_I$ is problematic seems borne out in many string theory models;
an example is D-brane models \cite{Baumann}. 

Very low $E_I$ has been argued to be theoretically problematic also.
Naive consideration of families of potential energy functions suggests that $E_I\lesssim2\times 10^{16}$\,GeV ($r\lesssim 0.01$) is non-generic \cite{Steinhardt}.
One of the most striking successes of high $E_I$ potentials is that they can 
naturally predict $n_s\sim 0.96$, and generic low-energy potentials fail to make this prediction.
Low $E_I$ potentials have the slow-roll parameter $\epsilon$ exponentially small, 
so that the observation $n_s=1-6\epsilon+2\eta=0.960\pm0.013$ \cite{WMAP5} implies $\eta=-0.02\pm0.0065$, 
leaving us wondering why $\eta$ is so small when it could just as well have been of order $-1$.
This problem is not alleviated by anthropic considerations~\cite{TegmarkInfP}.  
By using the observed value of density fluctuations, and setting $\epsilon<10^{-4}$ as the boundary,
$E\lesssim6.7\times 10^{15}$\,GeV defines this problematic low-scale region.
These theoretical issues for inflation are indicated by the vertical regions in Fig.~\ref{Naive}.
Although there are inflation models in the literature at energy scales both above and below this naive window, the debate about whether they are generic continues.

With theory in limbo, we turn to observational guidance.
High $E_I$ implies a large amplitude for primordial gravitational waves (GWs).
$E_I>3.8\times10^{16}$\,GeV ($r>0.22$) is ruled out by WMAP5 plus BAO and SN data \cite{WMAP5}, as indicated by the orange region of Fig.~\ref{ConsPlot}.  Possible future searches for primordial GWs have rightly been a focus of attention.  In this article we emphasize the 
additional information that can be learned from isocurvature fluctuations.

\subsection{Axion Physics}\label{AxionCos}

The QCD Lagrangian accommodates a gauge invariant, Lorentz invariant, renormalizable term $\propto \theta\,{\bf E}^a\!\cdot\!{\bf B}^a$, with $\theta\!\in\![-\pi,\pi]$, that manifestly breaks P and T symmetry.  Precision bounds on the electric dipole moment of the neutron constrain $|\theta|\lesssim10^{-10}$. The striking smallness of this parameter, which the standard model leaves unexplained, defines  the strong P and T problem (a.k.a.~``CP problem"). After introducing a new asymptotic (or alternatively, classical) Peccei-Quinn (PQ) symmetry \cite{PQ}
which is spontaneously broken, the effective $\theta$ becomes a dynamical variable, and relaxes toward extremely small values.  The consequent approximate Nambu-Goldstone boson is the axion 
\cite{WeinbergAxion,WilczekAxion}. 

\begin{figure}
\includegraphics[width=\columnwidth]{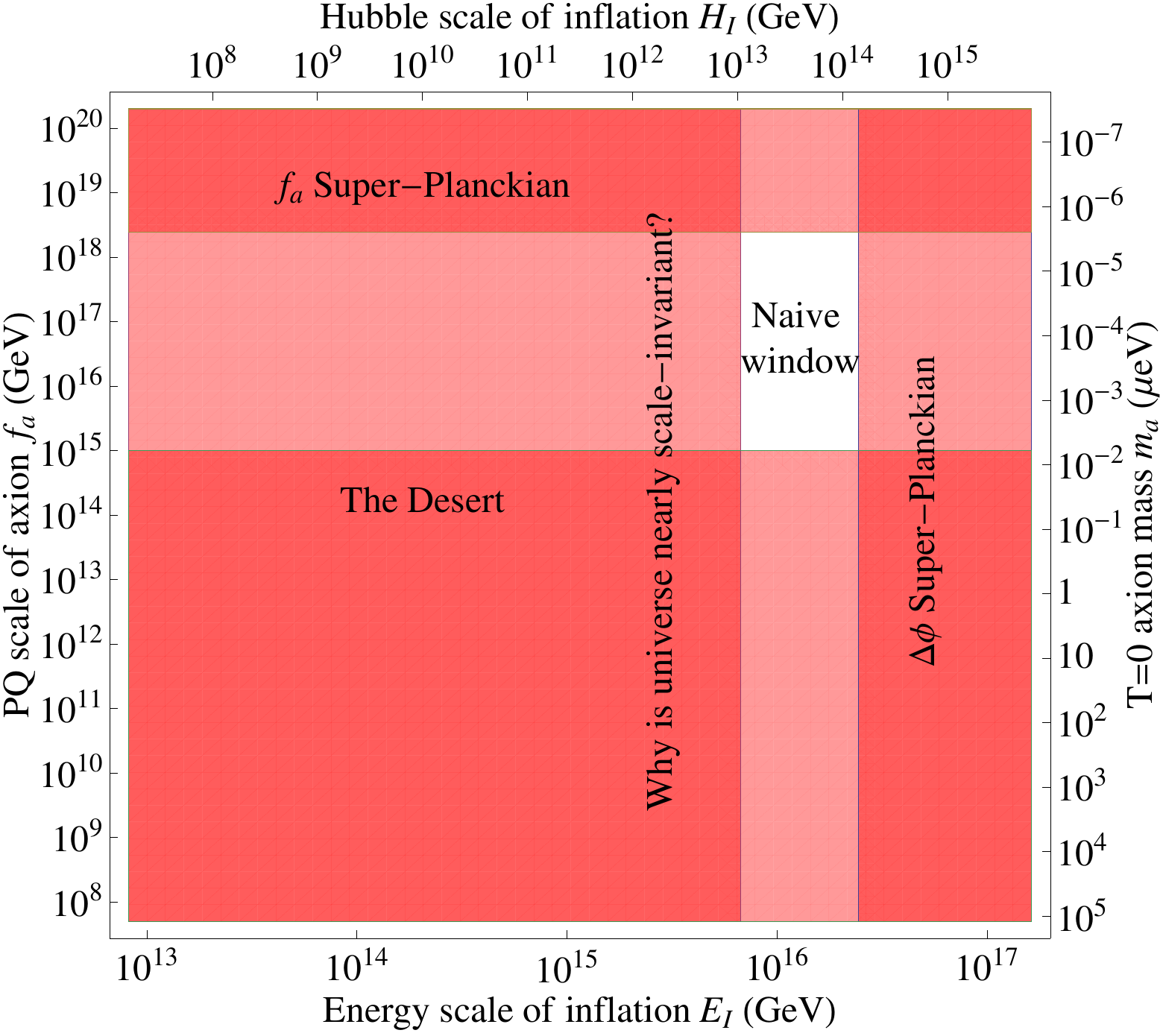}
\caption{{\em Naive expectations} for the energy scale of inflation $E_I$ 
and the axion PQ scale $\faN$. 
For $E_I\gtrsim2.4\times10^{16}$\,GeV, the inflaton must undergo super-Planckian excursions in field space (in single field models).
For $E_I\lesssim6.7\times10^{15}$\,GeV, generic inflation potentials fail to reproduce the observed nearly scale
invariant power spectrum.
For $f_a\gtrsim2.4\times10^{18}$\,GeV, the PQ breaking scale is super-Planckian.
For $f_a\lesssim10^{15}$\,GeV (and $f_a\gg$\,TeV), the PQ breaking is in the ``desert" of particle physics and non-trivial to achieve in string theory.
This leaves the region labeled ``naive window".}
\label{Naive}\end{figure}

\begin{figure}
\includegraphics[width=\columnwidth]{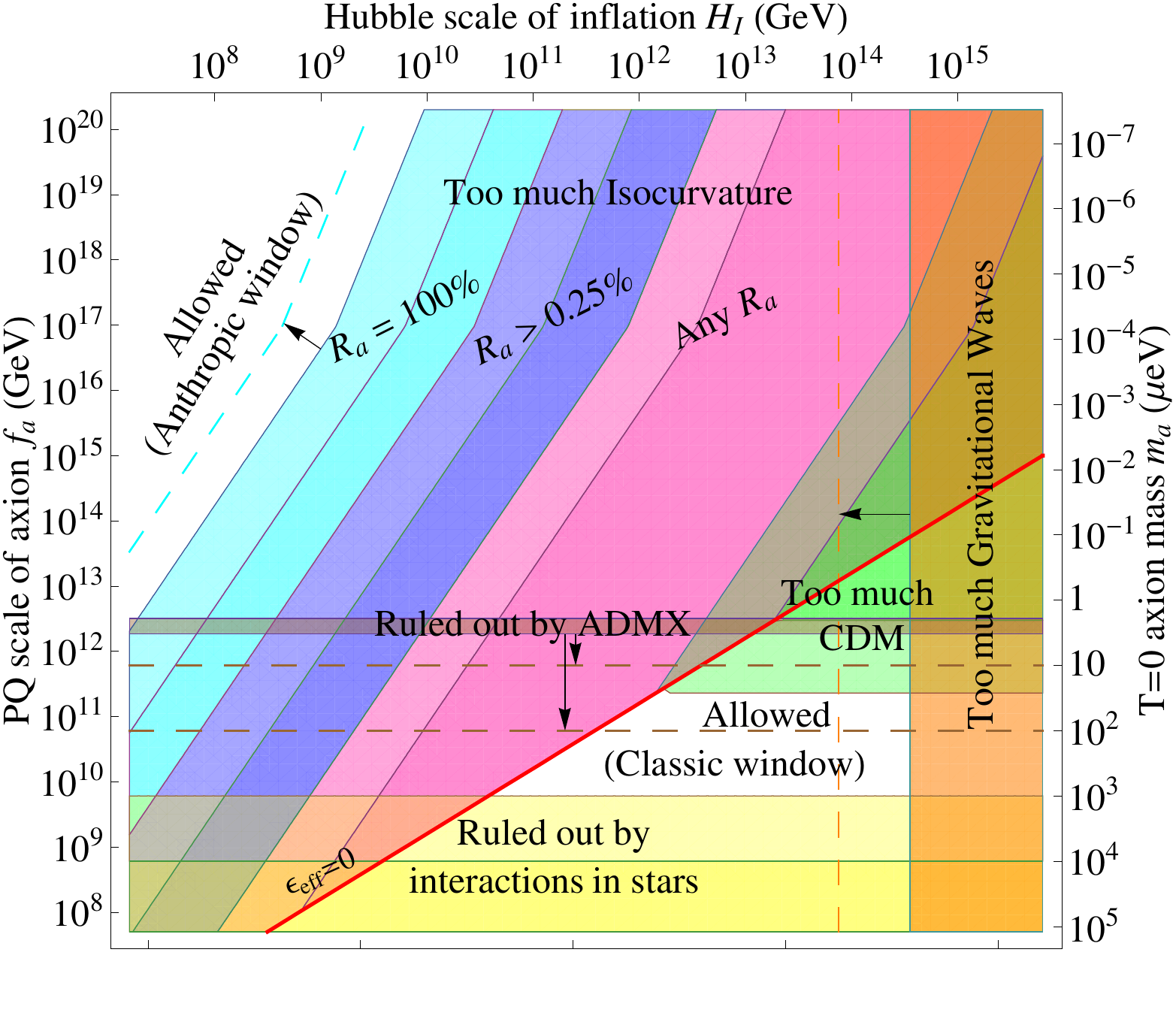}\vspace{-1.07cm}
\includegraphics[width=\columnwidth]{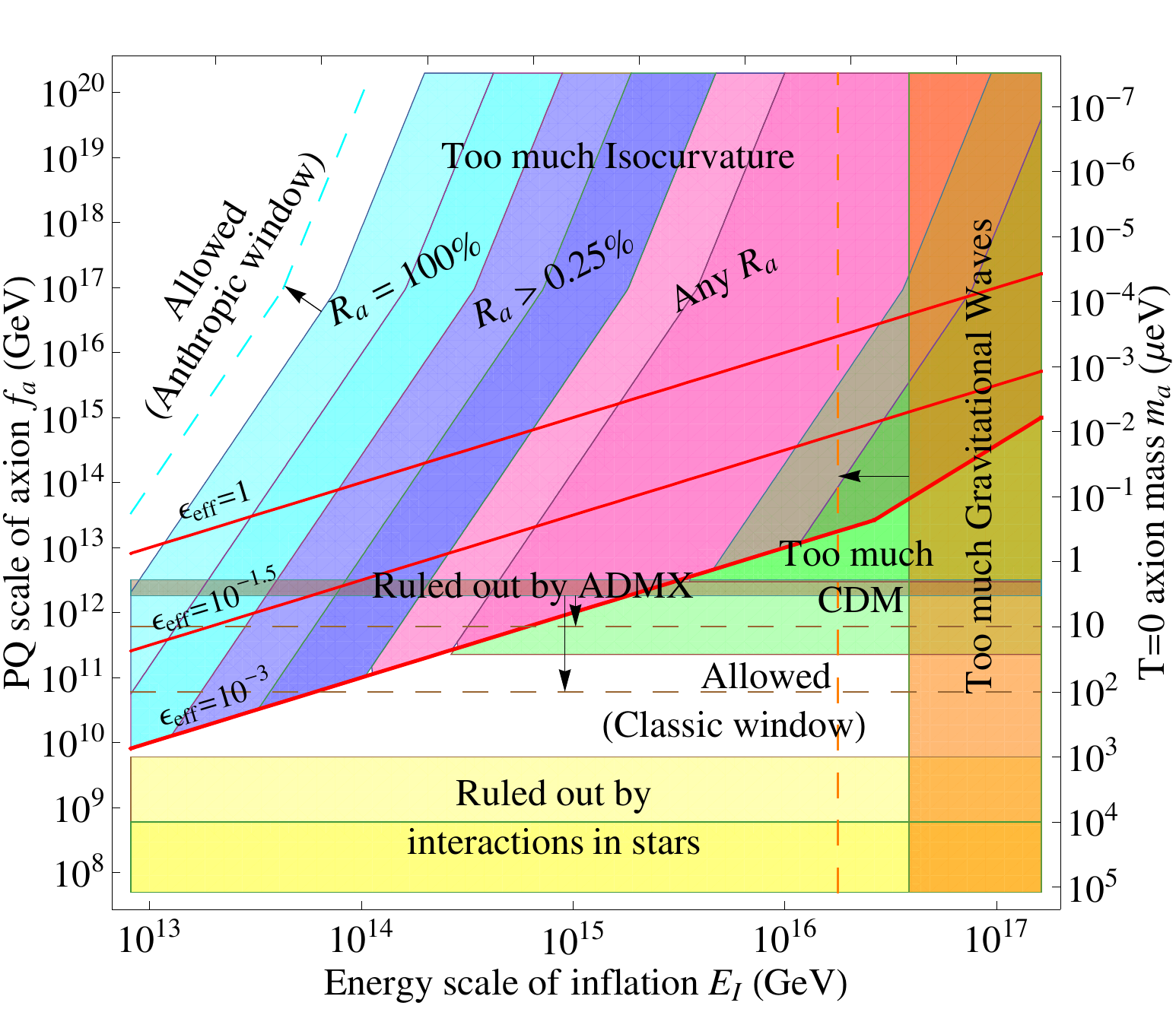}
\caption{{\em Observational constraints} on the energy scale of inflation $E_I$ 
and the axion PQ scale $\faN$ are shown in the top (bottom) panel for inefficient (efficient) thermalization 
at the end of inflation. 
The thick red diagonal line is $\faN=\mbox{Max}\{\TGH=H_I/2\pi,\TRH=\eff\,E_I\}$, ($\eff\approx0$ in top,
$\eff=10^{-3}$ in bottom, with $\eff=10^{-1.5}$, 1 indicated).
Above this line is the inflationary anthropic scenario and below this line is the classic scenario.
The region in which there is too much isocurvature, $\alpha_a>0.072$, depends on the axion fraction $R_a\equiv\xi_a/\cdm$ of the CDM;
the purple region applies for any $R_a$,
the blue region is for $R_a>0.25\%$ (which is expected at 95\% confidence),
and the cyan region is for $R_a=100\%$.
The green region has too much axion CDM: $\xi_a>2.9$\,eV.
Each constraint is divided into two parts: the darker part is for a conservative value $\tc=1/20$ and the lighter part is for a moderate value $\tc=1$.
The orange region has excessive GWs amplitude: $Q_t>9.3\times 10^{-6}$.
The yellow region has too much axion interaction in stars
(darker is firmly ruled out, lighter is for some analyzes).
The brown region is excluded by the laboratory ADMX search.
The dashed cyan, orange, and brown lines are future targets for isocurvature, GWs, and ADMX searches, respectively.}
\label{ConsPlot}\end{figure}

The simplest axion models contain only one phenomenologically significant parameter: $f_a$, the scale at which the PQ symmetry breaks.   The zero temperature Lagrangian for the complex field $\phi=\rho\,e^{i\theta}/\sqrt{2}$ is
\begin{equation}
\mathcal{L}=\frac{1}{2}f_a^2(\partial_\mu\theta)^2+\frac{1}{2}(\partial_\mu\rho)^2-2\Lambda^4\sin^2(\theta/2)-\lambda(|\phi|^2-f_a^2/2)^2\nonumber
\end{equation}
($\Lambda\approx 78$\,MeV, $\rho$ is irrelevant at low energies).
Accelerator bounds require $f_a$ to be well above the electroweak scale, and stellar astrophysics constraints place considerably higher limits.  
Given that electroweak values for $f_a$ are ruled out, economy suggests that $f_a$ could be associated with unification or Planck scales, rather than the \
``desert" of particle physics or super-Planckian scales, as indicated by the horizontal regions in Fig.~\ref{Naive}.  This intuition for $f_a$ seems borne out in string theory, where $f_a$ typically lies at or just above the GUT scale, and much lower values are non-trivial to 
achieve \cite{WittenAxion}.\footnote{For example, in weakly coupled heterotic string theory, the model-independent axion has its PQ scale given by $\faN=\alpha_U\mpl/(2\pi\sqrt{2})$.
A unified coupling $\alpha_U=1/25$ then gives $\faN\approx 1.1\times 10^{16}$\,GeV
\cite{Fox,WittenAxion}.} 
Such high values of $f_a$ correspond to large contributions from axions to cold dark matter (CDM).  Indeed, it is only after selection effects are taken into account that the ratio of axion density to entropy is small enough to be consistent with observations  \cite{LindeSelection, WilczekSelection}.  When these effects are included, one finds that the expected density of dark matter in axions is close to the amount of dark matter actually observed \cite{TegmarkSelection}.

\subsection{Cosmological Observables, Summary}

Quantum fluctuations in an effective inflaton field give rise to the standard adiabatic
fluctuations that have grown into our cosmologically observed large-scale structure. If the PQ symmetry undergoes spontaneous symmetry breaking before the end of inflation, 
quantum fluctuations in the consequent light axion field give rise to  
isocurvature fluctuations.
The amplitude of the isocurvature fluctuations grows with $E_I$, so upper bounds on the amplitude of isocurvature fluctuations 
imply upper bounds on $E_I$.

The purpose of this article is to delineate these bounds, extending earlier work on this subject such as \cite{Fox,Beltran,Burns,Lyth,LythStewart,TW}.
The bounds depend sensitively on the fraction $R_a$ of CDM in the form of axions, which in turn depends not only on $f_a$, but also on the local initial misalignment angle $\theta_i\!\in\![-\pi,\pi]$.  
These constraints are shown in Figs.~\ref{ConsPlot}, for different choices of 
the axion CDM fraction. We will estimate this fraction using statistical arguments.

In Section \ref{Iso} we calculate the production and late time abundance $\xi_a$ of axions and 
the amplitude $\alpha_a$ of isocurvature fluctuations, as well as reviewing
the amplitude $Q_t$ of primordial GWs.
These three observables depend on two micro-physical parameters:  the PQ scale $f_a$ (or equivalently, the $T=0$ axion mass $m_a$) and the energy scale of inflation $E_I$ (or equivalently, the Hubble scale of inflation $H_I$), and on one ``environmental'' parameter: the misalignment angle $\theta_i$. 
We summarize the final formulae here:
\begin{eqnarray}
\xi_a \amp = \amp
\M(\theta_i^2+\sigma_\theta^2)f(\theta_i^2)\tc F \label{abundeqn},\\
\alpha_a \amp = \amp
\frac{8}{25}\frac{(\M/\xi_m)^2}{{\langle(\delta T/T)^2_{\mbox{\tiny{tot}}}\rangle}} \sigma_\theta^2
(2\theta_i^2+\sigma_\theta^2) f(\theta_i^2)^2  \tc^2 F^2 \label{isoeqn},\\
Q_t  \amp = \amp \frac{H_I}{5\pi\,\mpl}=\frac{E_I^2}{5\sqrt{3}\,\pi\,\mpl^2},
\label{graveqn}\end{eqnarray}
where
\begin{eqnarray}
F  \amp\approx\amp \left\{
\begin{array}{ll}
2.8\,\left(\frac{\M}{\Lqcd}\right)^{2/3}
\left(\frac{\faN}{\mpl}\right)^{7/6}
\,\,\,\,\,\,\, \mbox{for}\,\,\,\faN \lesssim \fh,\\
4.4\,\left(\frac{\faN}{\mpl}\right)^{3/2}
\,\,\,\,\,\,\,\,\,\,\,\,\,\,\,\,\,\,\,\,\,\,\,\,\,\,\,\,\,\, \mbox{for}\,\,\,\faN \gtrsim \fh,
\end{array} \right. \\
\sigma_{\theta} \amp = \amp \gamma\frac{H_I}{2\pi f_a} = \gamma\frac{E_I^2}{2\sqrt{3}\,\pi f_a\mpl}
\label{defs}\end{eqnarray}
(definitions are given below).

The most recent observational bounds from WMAP5 combined with other data are \cite{WMAP5}
\begin{equation}
\xi_a\leq 2.9\,\mbox{eV},\,\,\,\,\alpha_a<0.072,\,\,\,\,Q_t<9.3\times 10^{-6},
\end{equation}
thereby constraining the two micro-physical parameters $E_I$ and $f_a$.

These expressions for $\xi_a$ and $\alpha_a$ only apply if the PQ symmetry undergoes spontaneous symmetry breaking before inflation and is not restored thereafter. 
This is true if $f_a$ exceeds the Gibbons-Hawking temperature during inflation and
the maximum post-inflationary thermalization temperature, as we discuss in Section \ref{Fluct}. 
Inefficient thermalization leads to constraints displayed in  Fig.~\ref{ConsPlot} (top), while efficient thermalization leads to constraints displayed in Fig.~\ref{ConsPlot} (bottom). 

In Section \ref{Dist}, we present a statistical estimate for the axion abundance 
to place additional constraints on out parameter space.
In Section \ref{Disc}, we conclude by discussing the implications for inflationary model building and future prospects.

\section{Axion Cosmology}\label{Iso}

In this section, we review the production of axions in the early universe, their abundance in the late universe, and the amplitude of isocurvature fluctuations, following Refs.~\cite{Burns,Fox},
and explain and derive eqs.~(\ref{abundeqn})--(\ref{defs}).
We focus on axion production from the so called ``vacuum misalignment" mechanism only. This provides the most conservative constraints.  Additional production mechanisms,
such as cosmic string decay, are subject to larger theoretical uncertainties (e.g., see \cite{Sikivie}).

\subsection{Onset of Axion Production}

\begin{figure}[t]
\includegraphics[width=\columnwidth]{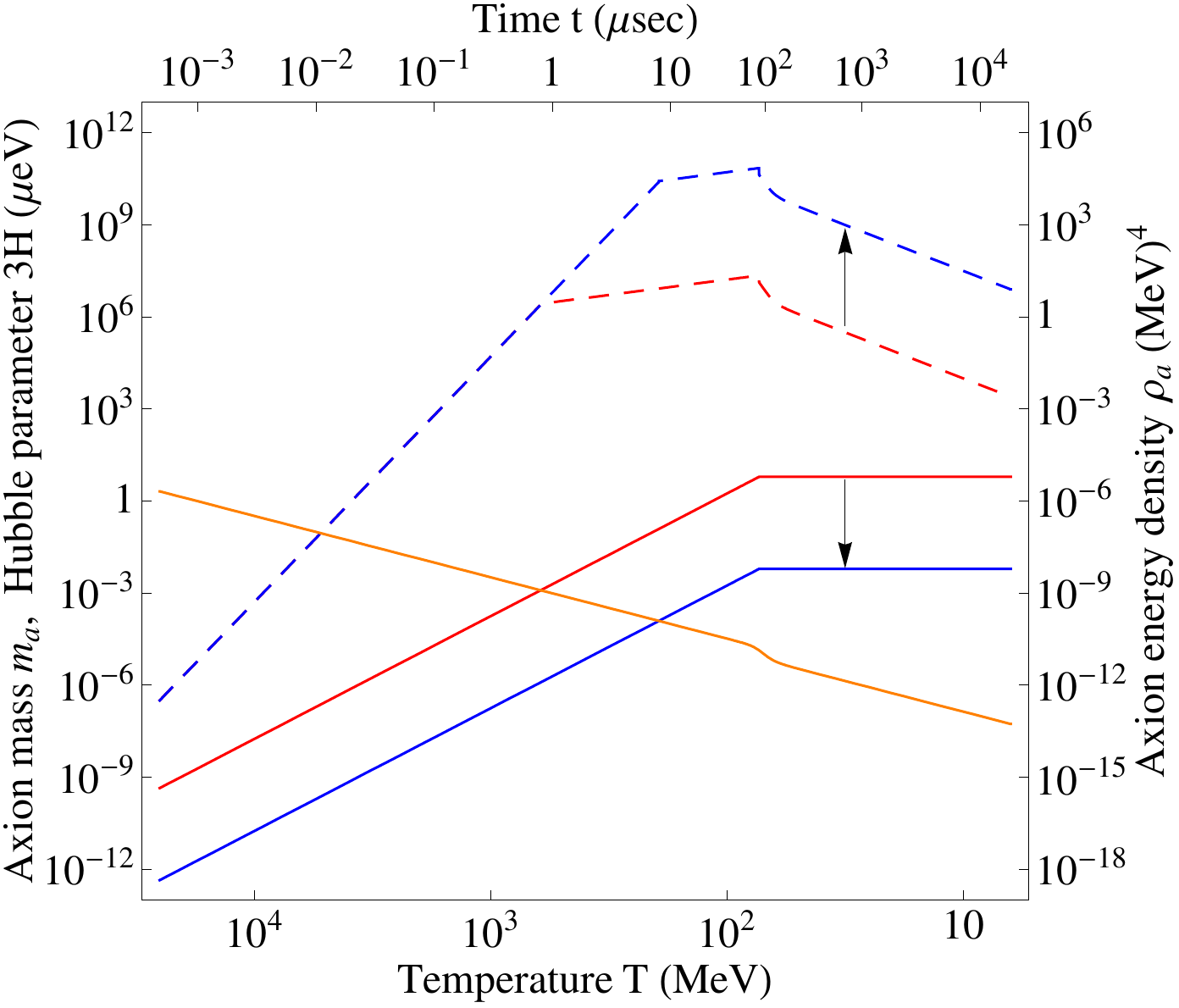}
\caption{
Axion mass $m_a$ (solid) and axion energy density $\rho_a$ (dashed), as a function of temperature $T$ (or time $t$). 
The red and blue curves are for PQ breaking scales $f_a=10^{12}$\,GeV and $f_a=10^{15}$\,GeV, respectively.
The arrow indicates the effect of increasing $f_a$. The decreasing orange curve is the Hubble parameter $3H$. 
We have taken $\langle\theta^2\rangle=\pi^2/3$, although this is modified by the intervention of inflation, as explained in the text. 
For simplicity, we have here only kept track of the variation with temperature in the number of relativistic degrees of freedom
before ($g_*=g_{*S}=61.75$) and after ($g_*=g_{*S}=10.75$) the QCD phase transition, and taken $\tc=1$.}
\label{Temp}\end{figure}

In an expanding flat FRW background at temperature $T$ with Hubble parameter $H(T)$, the phase field $\theta$ of broken PQ symmetry satisfies the equation of motion
\begin{equation}
\ddot{\theta}+3H(T)\dot{\theta}-\frac{\nabla^2\theta}{a^2}=-\frac{1}{f_a^2}\frac{\partial }{\partial\theta}V(\theta,T),
\label{aeom}\end{equation}
where dots indicate derivatives with respect to co-ordinate time.
Here $V(\theta,T)$ is the temperature dependent potential induced by QCD instantons. At zero temperature, $V(\theta,0)=\M^4(1-\cos\theta)$,
where $\M\approx 78$\,MeV sets the scale of the vacuum energy of QCD.\footnote{$\Lambda$ is set by $\Lqcd$ and quark masses: $\Lambda^2=\frac{\sqrt{z}}{1+z}f_\pi m_\pi$, $z\equiv m_u/m_d\approx 0.56$.}
For small values of the axion field, the potential is approximately harmonic: 
\begin{equation}
V(\theta,T)\approx \frac{1}{2}m_a(T)^2f_a^2\,\theta^2.
\end{equation} 
The mass is temperature dependent, with high and low $T$ limits given by
\begin{equation}
m_a(T) \approx m_a(0)\left\{
\begin{array}{ll}
b \left(\frac{\Lqcd}{T}\right)^4 & \mbox{for}\,\,\,T \gtrsim \Lqcd, \\
1 & \mbox{for}\,\,\, T \lesssim \Lqcd, \\
\end{array} \right.
\label{mT}\end{equation}
where $\Lqcd\sim 200$\,MeV is the scale at which QCD becomes strongly coupled, 
$b=\mathcal{O}(10^{-2})$ depends on detailed QCD physics, and $m_a(0)$ is
the zero temperature axion mass, related to the 
PQ scale $f_a$ and $\M$ by $m_a(0)=\M^2/\faN$.\footnote{If there are $N$ distinct vacua for $\theta$, 
then we should replace $f_a$ by $f_a/N$ here and throughout the article. 
However, any $N>1$ models are expected to have a large overabundance of energy density from domain walls, unless inflation intervenes.\label{ColorAnomaly}} The temperature dependence is illustrated in Fig.~\ref{Temp}.

In the early universe, the axion field is effectively massless, and so the right hand side of eq.~(\ref{aeom}) is negligible. Hence the zero mode of the axion field is essentially frozen due to Hubble friction.
When the temperature $T$ drops below $\Tosc$, defined by
\begin{equation}
3H(\Tosc)\approx m_a(\Tosc),
\end{equation}
the axion field will begin to oscillate, producing axions.  
Since this occurs during the radiation dominated era, we have\footnote{We assume here that the universe before BBN is adequately described by conventional physics. See Ref.~\cite{GrinNT} for other scenarios.}
\begin{equation}
H(T)^2=\frac{1}{3\mpl^2}\frac{\pi^2}{30}g_{*}(T)\,T^4,
\label{Friedmann}\end{equation}
where $\mpl\approx 2.4\times 10^{18}$\,GeV and $g_{*}(T)$, the effective number of relativistic degrees of freedom, depends on whether $\Tosc$ occurs before or after the QCD phase transition:
$g_{*}(\Tosc)=61.75$ for $\Tosc\gtrsim\Lqcd$ and $g_{*}(\Tosc)=10.75$ for $\Tosc\lesssim\Lqcd$.
Eqs.~(\ref{mT}--\ref{Friedmann}) allow us to solve for $\Tosc$
in terms of $\faN$:
\begin{equation}
\Tosc \approx \left\{
\begin{array}{ll}
0.36\,
\Lqcd^{2/3}\M^{1/3}\left(\frac{\mpl}{\faN}\right)^{1/6}
\,\, \mbox{for}\,\,\,f_a \lesssim \fh,\\
0.55\,
\M \left(\frac{\mpl}{\faN}\right)^{1/2}
\,\,\,\,\,\,\,\,\,\,\,\,\,\,\,\,\,\,\,\,\, \mbox{for}\,\,\,f_a \gtrsim \fh,
\end{array} \right.
\end{equation}
where $\hat{f}_a$, which reflects the break in eq.~(\ref{mT}), is defined in eq.~(\ref{match}) below.

\subsection{Density of Axions}

At the onset of production (when $T=\Tosc$) the axion energy density is
\begin{equation}
\rho_a(\Tosc)\approx \frac{1}{2}\,m_a(\Tosc)^2\faN^2\langle\theta^2\rangle f(\theta_i^2)\tc.
\label{prod}\end{equation}
Here $\langle\theta^2\rangle$ is the spatial average over our Hubble volume of the square of the initial misalignment. In terms of its mean $\theta_i$ and standard deviation $\sigma_\theta$,  $\langle\theta^2\rangle = \theta_i^2+\sigma_\theta^2$.
If the axion field is established before (or during the early stages of) inflation, then spatial variations in 
$\theta$ are smoothed out over our Hubble volume ($\nabla^2\theta/a^2\to0$).  Then $\theta_i=\langle\theta\rangle$ in our Hubble volume is an angle drawn from a 
uniform distribution: $\theta_i\!\in\![-\pi,\pi]$, with a small variance
that we discuss in the next subsection. On the other hand, if the axion field is established after inflation,
then $\theta_i=\langle\theta\rangle=0$, with variance $\sigma_\theta^2=\pi^2/3$ due to small scale variations.
$f(\theta_i^2)$ is a fudge factor acknowledging anharmonicity in the axion potential; for $\theta_i \to 0$,
$f(\theta_i^2)\to 1$.   
Finally, $\tc$ is a dimensionless correction factor due to temperature dependence during formation. In our numerics, we take $\Lqcd=200\,$MeV, $b=0.018$, and absorb all theoretical uncertainties into $\tc$. A conservative value is $\tc=1/20$ and a more moderate value is $\tc=1;$\footnote{See Refs.~\cite{Kim1,Kim2} for precise estimates of axion abundance.} both values are reported in Fig.~2.

By following the redshift as the universe expands from axion-formation to today, we can convert this initial energy density into a prediction for the present density, as illustrated in Fig.~\ref{Temp}.   Since we are focusing on the zero-mode, the axions form a non-relativistic Bose-Einstein condensate.  
At late times (say today's temperature $T_0$), the axion energy per photon is 
\begin{equation}
\xi_a=\frac{\rho_a(T_0)}{n_\gamma(T_0)} = 
\frac{m_a(T_0)}{m_a(\Tosc)}\frac{\rho_a(\Tosc)}{n_\gamma(T_0)} \frac{s(T_0)}{s(\Tosc)},
\end{equation}
where we have exploited the fact that entropy density $s(T)$ 
in a comoving volume is conserved.  The entropy density is given by
\begin{eqnarray}
s(T)=\frac{2\pi^2}{45}g_{*s}(T)\,T^3,
\end{eqnarray}
where $g_{*s}(T)$ is the effective number of relativistic degrees of freedom for $s(T)$ 
\cite{KolbTurnerBook}. 
Note that $g_{*s}(\Tosc)= g_{*}(\Tosc)$, $g_{*s}(T_0)$ comes from photons and neutrinos: $g_{*s}(T_0)=2+\frac{7}{8}\times 6\times\frac{4}{11}=3.91$, and $m_a(T_0)\approx m_a(0)$.
Since the number density of photons $n_\gamma(T_0) = 2\zeta(3)T_0^3/\pi^2$ depends on temperature in the same way that the late time axion energy density $\rho_a(T_0)$ does, the quantity $\xi_a$ is a temperature-independent, or equivalently time-independent, measure of the axion abundance.  
In contrast, the commonly used quantities $\Omega_a$ and $h^2\Omega_a$ do not tell us anything fundamental about our universe, since like $T$, they are effectively alternative time variables that 
evolve as our universe expands.
These different measures of axion density are related by $\Omega_a h^2\approx 0.0019\,(\xi_a/\mbox{1 eV})(T_0/\mbox{1 K})^3$, which at the present epoch
($T_0=2.725$\,K) reduces to $\Omega_a h^2\approx \xi_a/\mbox{26\,eV}$
($h$ is the dimensionless Hubble parameter).

Combining all this information yields eq.~(\ref{abundeqn}).
Note that higher $f_a$ (for a fixed value of $\langle\theta^2\rangle$) corresponds to higher axion energy density, as seen in Fig.~\ref{Temp}. 
The reason for this is that  higher $f_a$ corresponds to lower $m_a(0)$, so that the onset of axion production, when $3H(T)$ has fallen to $m_a(T)$, occurs later.
Hence there is less redshifting of the axion energy density after production (furthermore,
$\rho_a(\Tosc)$ is higher if $\faN<\fh$).

We locate the boundary between the  low and high $\faN$ limits by equating the two expressions for $\xi_a$.  They match when $\faN= \fh$, where
\begin{equation}
\fh\equiv 0.26\,\left(\frac{\Lambda}{\Lqcd}\right)^2\mpl.
\label{match}\end{equation} 
The observed total density of cold dark matter in our universe $\cdm\approx 2.9$\,eV implies 
$\xi_a \leq 2.9\,\mbox{eV}$ \cite{WMAP5}.

\subsection{Fluctuations from Inflation}\label{Fluct}

During inflation, the universe undergoes an approximately de Sitter phase with 
Hubble parameter $H_I$.
Quantum fluctuations during this phase induce several kinds of cosmological fluctuations.
\begin{itemize}
\item Adiabatic density fluctuations are generated, with an approximately scale-invariant spectrum.  The measured amplitude is 
$Q\approx 1.98\times 10^{-5}$ \cite{WMAP5}.\footnote{Here $Q=\frac{2}{5}\Delta_R(k=0.002\,\mbox{Mpc}^{-1})$ of Ref.~\cite{WMAP5}.}
\item Primordial gravity waves are generated, with an approximately scale-invariant spectrum whose amplitude is given in eq.~(\ref{graveqn}).
 WMAP5 plus BAO and SN data imply the bound: $Q_t<9.3\times 10^{-6}$ (95\%).
Since $Q$ is measured and $Q_t$ is set by $E_I$, 
the tensor to scalar ratio $r\equiv (Q_t/Q)^2$ is often used to characterize
the scale of inflation.  Using the same notation as \cite{WMAP5}, it is bounded by $r<0.22$ (95\%). 
\item Any other light scalar fields, such as the axion, are imprinted with fluctuations during inflation, similarly to gravitons.
The power spectrum of a canonically normalized scalar field $\phi$, such as $\phi_a$,  
in de Sitter space has a scale-invariant spectrum (e.g., see \cite{LindeReview})
\begin{equation}
\langle |\delta\phi_a(k)|^2\rangle=\left(\frac{H_I}{2\pi}\right)^2\frac{1}{k^3/2\pi^2}.
\end{equation}

It is essentially a thermal spectrum at  Gibbons-Hawking temperature $\TGH=H_I/2\pi$.
Fluctuations in the misalignment angle in $k$-space are scaled as $\sigma_\theta=\sigma_a/f_a$, per eq.~(\ref{defs}).
We write the corresponding fluctuations in real space as $\sigma_a = \gamma\,H_I/2\pi$, where $\gamma=\mathcal{O}(1)$ is a dimensionless constant.
Ref.~\cite{WMAP5} effectively takes $\gamma=1$ and Ref.~\cite{Lyth} argues that observations are sensitive to length scales corresponding to $\gamma \approx 4$,
while in our figures we have taken a moderate value of $\gamma=2$. These fluctuations provide a lower bound on $\xi_a$ and, as we discuss in the next subsection, on
isocurvature fluctuations.
\end{itemize}

In the preceding discussion, we assumed the existence of a light axion field during inflation.  
This is true only if PQ symmetry is broken before inflation. 
Furthermore, if PQ symmetry is restored after inflation, the fluctuations will be washed out.  PQ symmetry can be restored either by  the Gibbons-Hawking temperature during inflation, or by the maximum
thermalization temperature after inflation $\TRH$.\footnote{The maximum thermalization temperature should not be confused with the reheating temperature, 
which can be somewhat lower \cite{Kolb}. The maximum thermalization temperature is the maximum temperature of the thermal bath post-inflation, 
while the reheating temperature is the temperature at the end of the reheating phase, i.e., at  the beginning of the radiation era.} 
To characterize the maximum thermalization temperature,
we use a dimensionless efficiency parameter $\eff$ defined as
$\TRH = \eff\,E_I$, with $0<\eff<1$, with $\eff\ll 1$ expected.  

A robust criterion for the presence of the axion during inflation with fluctuations that survive is 
\begin{equation}
f_a>\mbox{Max}\{\TGH,\TRH\}. 
\end{equation}
If this condition is satisfied, inflationary expansion implies that $\theta_i\!\in\![-\pi,\pi]$ is drawn from a uniform distribution.
By postulating that $\theta_i$ is atypically small in our neighborhood (i.e., in our Hubble volume) one can accommodate large $f_a$.   This  defines what we term the anthropic regime 
(see Fig.~\ref{ConsPlot}).

Alternatively, if  $f_a<\mbox{Max}\{\TGH,\TRH\}$, then either there is no axion during inflation or its effects are washed out after inflation.  In this case $\theta^2$ fluctuates throughout our observable universe, 
with variance $\pi^2/3$, and there are no appreciable axion-induced isocurvature fluctuations. 
This defines what we term the classic regime (see Fig.~\ref{ConsPlot}).

\subsection{Isocurvature Fluctuations}

Fluctuations in the local equation of state $\delta(n_i/s)\ne 0$ at fixed total energy density $\delta\rho=0$ are known as isocurvature  fluctuations.  (In contrast, fluctuations with $\delta(n_i/s)=0$ and $\delta\rho\ne 0$ are known as adiabatic fluctuations.)
Since the axion is essentially massless in the early universe, at temperatures much greater than the QCD phase transition ($T\gg\Lqcd$), its energy density is entirely negligible at early times.
Hence, at such early times, fluctuations in the number density of axions (established by de Sitter fluctuations during inflation) do not alter the energy density of the universe.
Later, for temperatures below the QCD phase transition ($T\lesssim\Lqcd$),
the axion acquires a mass and a significant energy density (see Fig.~\ref{Temp}), but any such fluctuations cannot
alter the total energy density of the universe, by local conservation of energy. In the early radiation dominated era, this means that fluctuations in the axion energy density are compensated by fluctuations in photons and other relativistic fields. Hence, these are isocurvature 
fluctuations.\footnote{Later, around the onset of the matter dominated era, these isocurvature fluctuations are converted to adiabatic fluctuations, 
responsible for the familiar gravitational structures in our universe.}

To quantify the amplitude of isocurvature fluctuations, it is useful to introduce the fractional change in the number density to entropy density ratio:
\begin{equation}
S_i=\frac{\delta(n_i/s)}{n_i/s}=\frac{\delta n_i}{n_i}-3\frac{\delta T}{T}.
\end{equation}
For adiabatic fluctuations, $S_i=0$.  We assume that this is true for all species other than the axion. 
Isocurvature fluctuations in the total energy density involve a sum over all massive species and radiation:
\begin{equation}
0 = \delta\rhoi = m_a \delta n_a+\sum_{i\ne a}m_i \delta n_i+4\rho_r\frac{\delta T}{T}.
\label{sum}\end{equation}
These two equations will be used to obtain an expression for the 
corresponding temperature fluctuations.

Initially the energy density of the axion field is a small fraction of the ambient total density, 
so eq.~(\ref{sum}) gives $\delta T/T\ll\delta n_a/n_a$, and $S_a=\delta n_a/n_a$. Since $n_a\propto \theta^2$ (ignoring anharmonic effects), this implies
\begin{equation}
S_a = \frac{\theta^2-\langle\theta^2\rangle}{\langle\theta^2\rangle}.
\end{equation}
Assuming $\delta\theta\equiv\theta-\langle\theta\rangle$ is Gaussian distributed\footnote{This is a good assumption in the regime where the axions comprise a significant fraction of the dark matter, i.e., $\theta_i^2\gg\sigma_\theta^2$.}, we can calculate $\langle S_a^2\rangle$ in terms of 
$\theta_i=\langle\theta\rangle$ and $\sigma_\theta^2=\langle(\delta\theta)^2\rangle$, as
\begin{equation}
\langle S_a^2\rangle = \frac{2\sigma_\theta^2(2\theta_i^2+\sigma_\theta^2)}{(\theta_i^2+\sigma_\theta^2)^2}.
\end{equation}
Note that if $\theta_i^2\ll\sigma_\theta^2$ then $\langle S_a^2 \rangle = 2$,
while if $\theta_i^2\gg\sigma_\theta^2$ then $\langle S_a^2\rangle=4\sigma_\theta^2/\theta_i^2$.

The most important axion induced temperature fluctuations are those on the largest scales.
Such fluctuations enter the horizon well into the matter dominated era, where $\rho_r$ can be ignored.
This implies\footnote{Due to the Sachs-Wolfe effect, there is a 20\% 
enhancement to (\ref{horizon}), but we will not go into those details here.}
\begin{equation}
\left(\frac{\delta T}{T}\right)_{\mbox{\tiny{iso}}}\approx-\frac{\xi_a}{3\,\xi_m}S_a,
\label{horizon}\end{equation}
where $\xi_m$ is the total matter energy density per photon, whose measured value is
$\xi_m = 3.5\,\mbox{eV}$ \cite{WMAP5}.

Following \cite{WMAP5}, we define $\alpha_a$ to be the fractional contribution to the CMB 
temperature power spectrum due to axion isocurvature:
\begin{eqnarray}
\alpha_a \amp \equiv \amp \frac{\langle(\delta T/T)^2_{\mbox{\tiny{iso}}}\rangle}
{\langle(\delta T/T)^2_{\mbox{\tiny{tot}}}\rangle}.
\end{eqnarray}
Using the relationship between $S_a$ and $\delta T/T$ in eq.~(\ref{horizon})
and the preceding expression for $\langle S_a^2\rangle$
we obtain
\begin{eqnarray}
\alpha_a \amp = \amp \frac{8}{25}\frac{(\xi_a/\xi_m)^2}
{\langle(\delta T/T)^2_{\mbox{\tiny{tot}}}\rangle} 
\frac{\sigma_\theta^2(2\theta_i^2+\sigma_\theta^2)}
{(\theta_i^2+\sigma_\theta^2)^2}.
\label{isopre}\end{eqnarray}
COBE measured (and WMAP confirmed) the root-mean-square total temperature fluctuations to be
${\langle(\delta T/T)^2_{\mbox{\tiny{tot}}}\rangle}\approx (1.1\times 10^{-5})^2$, averaged over the first few $l$.
Using the expression for $\xi_a$ given in eq.~(\ref{abundeqn}) 
we can write the isocurvature fluctuations as in eq.~(\ref{isoeqn}).
This must be consistent with the latest observational bound
$\alpha_a < 0.072$. Here we have used $\alpha_0$ of Ref.~\cite{WMAP5}, which assumes
isocurvature fluctuations are uncorrelated from curvature ones.

\section{Direct and Statistical Constraints}\label{Dist}

It is conceivable that the axion abundance is negligible (but see the following subsection).  
This scenario (case (i)) requires $\theta_i^2\ll\sigma_\theta^2$. By demanding $\alpha_a<0.072$ (the current isocurvature bound) and using eq.~(\ref{isoeqn}), we obtain the most conservative bound, studied in Ref.~\cite{Fox}, 
corresponding to the purple region marked ``Any $R_a$'' in Fig.~\ref{ConsPlot}.

At the other extreme, if axions are the dominant form of dark matter in the universe (case iii), then
$\theta_i^2\gg\sigma_\theta^2$. Again demanding $\alpha_a<0.072$ in eq.~(\ref{isoeqn}), 
with $\theta_i$  
determined from eq.~(\ref{abundeqn}) with $\xi_a=\cdm$, the excluded region expands to include the cyan region marked ``$R_a=100\%$" in Fig.~\ref{ConsPlot} 
(as well as the blue region marked ``$R_a>0.25\%$'').

Each of these three regions are bifurcated by a line. In all three cases, the rightmost part gives the most robust constraint, coming from a
conservative value $\tc=1/20$, while the leftmost part extend the constraints using a more moderate (and more speculative) value $\tc=1$. 
This comes from our uncertainty in the total axion abundance.

\subsection{Statistics of a Two-Component Model}

The viability of large $f_a$ axion cosmology depends on taking selection effects seriously, since they produce a higher dark matter density $\xi$ than observed in most Hubble volumes. 
In particular, the density of a typical galaxy scales as $\rho\sim\xi^4$. Taking into account that denser galaxies have fewer stable solar systems due to close encounters with other stars, etc., 
it has been found that typical stable solar systems in large $f_a$ axion models reside in Hubble volumes where $\xi$ is comparable to the observed value \cite{TegmarkSelection}.

Here we draw out a statistical implication for the predicted axion abundance, if there is a second contributor to the dark matter density. 
Consider the hypothesis that the total CDM ($\cdm$) is comprised of axions ($\xi_a$) 
and some other component, say WIMPs ($\xi_W$):
$\cdm=\xi_a+\xi_W$.   The unknown separate axion and WIMP abundances should be drawn from prior distributions determined by  underlying micro-physical theories.  For axions in the large $f_a$ regime, above any inflation temperatures, this scenario implies that the initial misalignment angle $\theta_i$ is uniformly distributed.
In the regime where the axion abundance is non-negligible ($\theta_i^2\gg\sigma_\theta^2$), but still sufficiently small that we can ignore anharmonic effects
($\theta_i^2\ll 1$), we have $\xi_a\propto \theta_i^2$.   
Since $\theta_i$ is uniformly distributed, it is simple to show that
\begin{equation}
p^{(a)}_{\mbox{\tiny{prior}}}(\xi_a)\propto\frac{1}{\sqrt{\xi_a}}.
\label{axionprior}\end{equation}
In contrast, we do not have a reliable prior distribution $p^{(W)}_{\mbox{\tiny{prior}}}(\xi_W)$ for the WIMP.   

As discussed in Ref.~\cite{TegmarkSelection}, selection effects depend 
only on the  sum $\cdm=\xi_a+\xi_W$,
so the total joint distribution for axions and WIMPs is
\begin{equation}
p(\xi_a,\xi_W)\propto p^{(a)}_{\mbox{\tiny{prior}}}(\xi_a)p^{(W)}_{\mbox{\tiny{prior}}}(\xi_W)
p_{\mbox{\tiny{selec}}}(\cdm)
\end{equation}
As demonstrated in Ref.~\cite{TegmarkSelection}, the observed value of CDM $\cdm\approx 2.9$\,eV
is nicely consistent with this distribution. Given this,
we can focus on the remaining one-dimensional distribution for the axion:
\begin{equation}
p(\xi_a)\propto p^{(a)}_{\mbox{\tiny{prior}}}(\xi_a)p^{(W)}_{\mbox{\tiny{prior}}}(\cdm-\xi_a).
\end{equation}
Unless $p^{(W)}_{\mbox{\tiny{prior}}}$ is sharply peaked at $\cdm$, the axion prior (when integrated) disfavors very small values of $\xi_a$.    For example, let us take the WIMP prior to be uniform. 
We can then make a prediction for the axion abundance at, say, 95\% confidence.
Defining $\hat{\xi}_a$ implicitly through 
\begin{equation}
\int_0^{\hat{\xi}_a} p(\xi_a)\,d\xi_a=0.05
\label{lowerbd}\end{equation}
and solving eq.~(\ref{lowerbd}) using eq.~(\ref{axionprior}), we find 
$\hat{\xi}_a=(0.05)^2\cdm=0.25\%\,\cdm$.
This says that it statistically unlikely -- at the 95\% level --  for axions to comprise less than 0.25\% of the
CDM of the universe (case(ii)).

By setting $\xi_a=0.25\%\,\cdm$, we rule out the blue region marked $R_a>0.25\%$ in Fig.~\ref{ConsPlot} with high confidence.
In other words, without assuming that axions comprise all the CDM, we find that on statistical grounds axions must comprise at least 
a non-negligible fraction of the universe's CDM, allowing us to extend the excluded region in  Fig.~\ref{ConsPlot} further towards the upper left.

\subsection{Additional Constraints}\label{Constraints}

The preceding applies in the 
$\faN>\mbox{Max}\{\TGH,\TRH\}$ regime, where the initial misalignment angle $\theta_i$ takes on a single constant value in our Hubble volume.
For $\faN<\mbox{Max}\{\TGH,\TRH\}$, the misalignment angle varies on cosmologically small scales, with average $\langle\theta^2\rangle=\pi^2/3$. 
In this regime the isocurvature fluctuations are negligible. In this case, bounds arise from the requirement that the axion abundance is not greater than the observed total CDM abundance: 
$\xi_a\leq\cdm\approx 2.9$\,eV. Using the upper expression for $\xi_a$ in eq.~(\ref{abundeqn}),
with $\theta_i^2+\sigma_\theta^2\to\langle\theta^2\rangle=\pi^2/3$, we find that
$f_a>2.3\times10^{11}\,\tc^{-6/7}$\,GeV is ruled out. 
For the conservative value $\tc=1/20$, this excludes the upper part of the green region marked ``Too much CDM'' in Fig.~\ref{ConsPlot},
and for the moderate value $\tc=1$ this extends the 
exclusion to the lower part of the green region.\footnote{If $f_a>\mbox{Max}\{\TGH,\TRH\}$, there 
is another region ruled out with too much CDM (green region above thick red line in Fig.~\ref{ConsPlot}.)} 

Also, $m_a(0)>10^3\,\mu$eV is firmly ruled out (and $m_a(0)>10^4\,\mu$eV for some analyzes), 
since in this regime the coupling of axions to matter is too large, affecting the physics of stars, 
such as the cooling of red giants and the neutrino flux from SN 1987A \cite{pdg} 
(yellow region at bottom of Fig.~\ref{ConsPlot}.) 
Furthermore, the ADMX search for axion dark matter in a microwave cavity detector
has ruled out axions comprising the bulk of the halo dark matter 
in the following mass window: $1.9\,\mu$eV$<m_a(0)<3.3\,\mu$eV (brown band in Fig.~\ref{ConsPlot}) 
for so-called KSVZ axions, and the sub-window $1.98\,\mu$eV$<m_a(0)<2.17\,\mu$eV for so-called DFSV axions \cite{ADMX1,ADMX2}.
The remaining white region is the allowed ``classic window".

In Fig.~\ref{ConsPlot} (top), corresponding to inefficient thermalization ($\eff\approx 0$), 
the boundary between the anthropic and classic regimes is $\faN=\TGH=H_I/2\pi$. 
In Fig.~\ref{ConsPlot} (bottom),  corresponding to efficient thermalization ($\eff=10^{-3}$, with $\eff=10^{-1.5}$, 1 indicated), the boundary between the two regimes is  $\faN=\TRH=\eff E_I$. 
Efficient thermalization thus opens up a larger ``classic window", but the ``anthropic window" is essentially unaltered. 

\subsection{Effect of Falling Density During Inflation}\label{Comments}

In our analysis, we have treated inflation as occurring at one rather well-defined Hubble scale. 
Although this is a good approximation in some inflation models, there are others giving an 
appreciable change in $H$ between its value (say $H_I$) when the modes that are now re-entering our
horizon left the horizon (55 or so e-foldings before the end of inflation), and its value (say $H_{\mbox{\tiny{end}}}$) at
the end of inflation. 
This is particularly relevant to Fig.~\ref{ConsPlot} (top), since it implies that the boundary between
``anthropic" and ``classic" regimes is blurred, since $\TGH=H/2\pi$ is evolving. For high scale inflation, $H$ typically changes by an amount
 of order the number of e-foldings, i.e., $\mathcal{O}(10^2)$, while for low scale inflation models, $H$ typically changes very little.

If we consider $H_{\mbox{\tiny{end}}}\ll H_I$, then the PQ symmetry can break {\em during} inflation. 
The resulting cosmology could be quite interesting with axion dark matter varying appreciably from one point in our Hubble volume to another, but is ruled out since $Q\sim 10^{-5}$.
If PQ breaking occurs
very close to 55 or so e-foldings before the end of inflation, then $\theta_i$ can be smoothed out on today's cosmological scales and make the analysis ``anthropic". Otherwise, we expect the ``classic" analysis to apply as usual, providing a ruled out green region in Fig.~\ref{ConsPlot}. Hence, we expect
such corrections to the constraints to be reasonably minimal. 

\section{Discussion}\label{Disc}

We have surveyed observational constraints on the parameter space $\{E_I, f_a\}$ of inflation and axions, finding that most of it is excluded, leaving 
only two allowed regions that we term classic and anthropic windows.
Part of the classic window $\faN\sim 10^{11}$ -- $10^{12}$\,GeV
will be intensely explored by the ongoing ADMX experiment. 
The region indicated by the arrow to the horizontal brown lines in Fig.~\ref{ConsPlot} to $m_a(0)=10\,\mu$eV is expected to be explored by the end of ADMX Phase II,
and onwards to $m_a(0)=100\,\mu$eV some years thereafter \cite{ADMXfuture}.
In this window, comparatively little can be concluded about the scale of inflation. From Fig.~\ref{ConsPlot} (top), taking $m_a(0)=100\,\mu$eV and assuming $\xi_a>0.25\%\,\cdm$, we rule out
$8.4\times 10^{13}\,$GeV $\lesssim E_I\lesssim 1.3\times 10^{15}\,$GeV.\footnote{The quoted lower end of the ruled out region is the geometric mean 
of the conservative and moderate $\xi_a$-scenarios.\label{Intermediate}}
From Fig.~\ref{ConsPlot} (bottom), the upper end of this ruled out region is reduced due to efficient post-inflation thermalization.
Although we can rule out a range of low scale inflation models,
these conclusions are not exceedingly strong.



On the other hand, a large $\faN$ axion has strong implications for inflation.
According to both Figs.~\ref{ConsPlot} (top) \& (bottom), 
if $f_a=10^{16}$\,GeV then 
$E_I\gtrsim5.5\times 10^{14}$\,GeV ($r\gtrsim9\times10^{-9}$) 
is ruled out at 95\% confidence for the conservative value $\tc=1/20$, and 
$E_I\gtrsim2.6\times 10^{14}$\,GeV ($r\gtrsim4\times10^{-10}$) 
is ruled out for the moderate value $\tc=1$. 
The geometric mean is $E_I\gtrsim3.8\times 10^{14}$\,GeV ($r\gtrsim2\times10^{-9}$),
which is reported in the abstract. 
This is incompatible with many models of inflation, including ``classic'' models with a single slow-rolling scalar field in a generic potential. For example, 
monomial potentials $V \propto \phi^p$ predict $r=4\,p/N_e$, where $N_e$ is the number of e-foldings of inflation from when it generated our horizon scale fluctuations to when it ended. 
For such models, $N_e$ around 50 or 60 is expected, so any reasonable $p$ is ruled out, including $\phi^2$ chaotic inflation \cite{Inf4}
and the stringy N-flation \cite{Nflation} and Monodromy \cite{Silverstein} models. 
The same is true for exponential potentials $V \propto \exp(-\sqrt{2p}\,\phi/\mpl)$, which predict $r=16p$. 

Evidently there is considerable tension between the theoretically appealing large $f_a$ and high-scale inflation scenario (see Fig.~\ref{Naive}) and the observational constraints (see Fig.~\ref{ConsPlot}).
Low-scale inflation may be emerging as favored from recent work in string theory. If we consider the small subspace (see \cite{Delicate,Hertzberg1,Hertzberg2}) of presently constructed string models that both inflate and agree with the observed values of $Q$ and $n_s$, 
we are left with models that tend to be at rather small energies, typically $r<10^{-8}$ for D-brane models 
and various other scenarios such as modular inflation \cite{Conlon}.
There are also arguments for very low $r$ in the simple KKLT framework discussed in Ref.~\cite{KalloshLinde}. 
This allows $f_a\sim 10^{16}$\,GeV to be marginally consistent with present isocurvature bounds.  Although it is highly premature to conclude that very low energy scale is a generic feature of
string realizations of inflation, it is intriguing that many string constructions have this feature.
(See \cite{Nflation,Silverstein} for interesting exceptions.)

The Planck satellite, CMBPol, and upcoming suborbital CMB experiments should probe well beyond the current 
bound on GWs of $r<0.22$, perhaps reaching $r\sim 0.01$. This is indicated by an arrow toward the vertical dashed red line in Fig.~\ref{ConsPlot}.
If gravity waves are observed in this regime, then the PQ scale $\faN$ must be in the classic window.

Our considerations emphasize the fundamental importance of improving  bounds on
isocurvature fluctuations. For example, an order of magnitude improvement to $\alpha\sim 0.007$
would push the isocurvature bounds to the diagonal dashed cyan line in Fig.~\ref{ConsPlot}.
(We have indicated the improvement for the case  where axions comprise all the CDM: $\xi_a=\cdm$.)
Detection of isocurvature fluctuations in this regime has three important implications:
\begin{enumerate} 
\item It could be interpreted as evidence for the existence of the axion field, and assuming this:
\item It would probe low inflation scales $E_I$ far beyond the scope of any foreseeable GW measurements. 
\item It would be evidence that we live in a highly atypical Hubble volume, i.e., $\{E_I,\faN\}$ must be in the anthropic window.
\end{enumerate}
Isocurvature modes and tensor modes thus provide complementary constraints on fundamental physics,  
making it fruitful to study dark matter and inflation in a unified way.

\acknowledgments
We would like to thank Daniel Baumann, Mauro Brigante, Andrea de Simone, Shamit Kachru, David B.~Kaplan, Tongyan Lin, Gray Rybka, and Paul Steinhardt, for
helpful discussions, and a special thanks to Krishna Rajagopal and Scott Watson for comments on a preliminary version of this paper.
M.~P.~H. and F.~W. thank the Department of Energy (D.O.E.) for support under cooperative research agreement DE-FC02-94ER40818.
This work was supported by
NSF grants AST-0134999 and AST-05-06556, 
a grant from the John Templeton foundation
and fellowships from the David and Lucile
Packard Foundation and the Research Corporation.  

\end{document}